# Complexity Reduction of Volterra Nonlinear Equalization for Optical Short-Reach IM/DD Systems

Tom Wettlin, Stephan Pachnicke
*Chair of Communications*
*Kiel University*
24143 Kiel, Germany
tom.wettlin@tf.uni-kiel.de

Talha Rahman, Jinlong Wei,
Stefano Calabrò, Nebojsa Stojanovic
*European Research Center*
*Huawei Technologies*
80992 Munich, Germany

*Abstract*—We investigate approaches to reduce the computational complexity of Volterra nonlinear equalizers (VNLEs) for short-reach optical transmission systems using intensity modulation and direct detection (IM/DD). In this contribution we focus on a structural reduction of the number of kernels, i.e. we define rules to decide which terms need to be implemented and which can be neglected before the kernels are calculated. This static complexity reduction is to be distinguished from other approaches like pruning or $L_1$ regularization, that are applied after the adaptation of the full Volterra equalizer e.g. by thresholding. We investigate the impact of the complexity reduction on 90 GBd PAM6 IM/DD experimental data acquired in a back-to-back setup as well as in case of transmission over 1 km SSMF. First, we show, that the third-order VNLE terms have a significant impact on the overall performance of the system and that a high number of coefficients is necessary for optimal performance. Afterwards, we show that restrictions, for example on the tap spacing among samples participating in the same kernel, can lead to an improved tradeoff between performance and complexity compared to a full third-order VNLE. We show an example, in which the number of third-order kernels is halved without any appreciable performance degradation.

## I. Introduction

The worldwide data traffic is continuously growing due to cloud computing, video streaming etc. A large part of this traffic occurs inside data centers (DCs). There high-speed connections between different servers are necessary. These short-reach connections of up to 2 km need to be realized in a cost-efficient manner. Therefore, simple intensity-modulated direct-detection (IM/DD) systems using only a single photodiode for signal reception are currently preferred over the more complex coherent systems. For these IM/DD systems, different modulation formats are currently investigated. Pulse-amplitude modulation (PAM) is often preferred over alternatives like discrete multitone (DMT) [1] for its performance and simplicity. For the next generation of intra data-center links, a data rate as high as 800 Gb/s is aspired. This rate will most likely be realized in a four-carrier CWDM system with 200 Gb/s/λ. This approach saves half the transceiver hardware compared to the realization using eight wavelengths with 100 Gb/s/λ each. However, a higher per lane data rate demands high symbol rates, which leads to severe impairments of the signal due to bandwidth limitations, chromatic dispersion and other effects.

To overcome these impairments, powerful digital signal processing (DSP) is essential. A popular approach to overcome both, linear and nonlinear impairments, is Volterra nonlinear equalization (VNLE). It has been applied in several publications on short-reach IM/DD systems [2-4]. For optical communications, a restriction to the third order is typical. Still, the complexity of VNLE can be very high and therefore be conflicting with the demand on low-complexity and low-cost systems. Different schemes to reduce the complexity of VNLEs were already investigated. However, many of these schemes like the $L_1$ regularization [5] or pruning [6] reduce the number of kernels after the adaptation of the full VNLE, which means that all kernels would still need to be considered in a practical implementation. For this reason, we focus on structural kernel reduction schemes, which define the considered kernels in advance of the actual adaptation. We investigate these reduction schemes based on data acquired in 90 GBd PAM6 and 75 GBd PAM8 experiments. The target of the structural kernel reduction is to improve the tradeoff between performance and complexity compared to the full VNLE.

The remainder of this paper will be structured as follows. In Section II the basics of VNLE for IM/DD systems are explained shortly. Afterwards, the investigated approaches for a structural complexity reduction are shown. The performance of the complexity reduction schemes is extensively investigated based on experimental data in Section III. Finally, Section IV concludes the paper.

## II. Volterra Nonlinear Equalization

VNLE is a widely spread approach to mitigate linear and nonlinear impairments in fiber optic communication systems. In short-reach links that we focus on, nonlinear impairments are mainly caused by the transceiver hardware, rather than by fiber nonlinearities. One source of nonlinearities is the nonlinear response of the modulator, e.g. the cosine characteristic of the Mach-Zehnder modulator (MZM). Also, the square-law detection of the photodiode (PD) distorts the signal in a nonlinear way. Additional nonlinearities can stem from amplifiers and lasers. To combat these impairments, a restriction of the VNLE to the third order is common, since the complexity rises significantly with higher orders. The third-order VNLE for IM/DD systems is defined as

$$y(k) = w_{dc} + \sum_{k_1=0}^{M_1-1} w_1(k_1)x(k-k_1)$$
$$+ \sum_{k_1=0}^{M_2-1}\sum_{k_2=k_1}^{M_2-1} w_2(k_1,k_2)x(k-k_1)x(k-k_2) \quad (1)$$
$$+ \sum_{k_1=0}^{M_3-1}\sum_{k_2=k_1}^{M_3-1}\sum_{k_3=k_2}^{M_3-1} w_3(k_1,k_2,k_3)x(k-k_1)x(k-k_2)x(k-k_3),$$

where $M_1$, $M_2$ and $M_3$ are the memory lengths for the linear terms and the nonlinear terms of second and third order. The matrices $w_1$, $w_2$ and $w_3$ contain the equalizer tap values, $w_{dc}$ is the coefficient that is responsible for the DC component of the sequence, and $x(k)$ is the input sequence. The relationship between the memory lengths and the number of equalizer kernels is shown in Tab. 1. The number of multiplications is 1 per linear kernel, 2 per second-order kernel and 3 per third-order kernel. Obviously, the third-order kernels have a big contribution to the overall computational complexity. Therefore, we focus on approaches to reduce this number in the following. To optimize the tap weights, we used training-symbol based least-mean squares (LMS) algorithm for coarse convergence and decision-directed (DD) LMS for fine adaptation. An extensive description of the VNLE and adaptation algorithms can be found in [7].

TABLE I
RELATIONSHIP BETWEEN MEMORY LENGTH AND KERNEL NUMBER FOR FIRST TO THIRD-ORDER VNLE

| VNLE Order | Number of Kernels |
| --- | --- |
| 1st | $N_1 = M_1$ |
| 2nd | $N_2 = M_2(M_2+1)/2$ |
| 3rd | $N_3 = M_3(M_3+1)(M_3+2)/6$ |

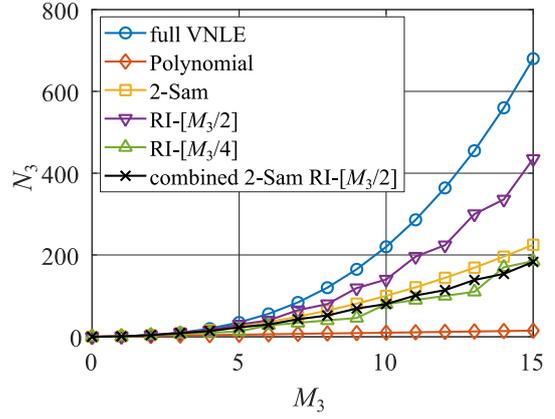

Fig. 1. Relation between the third-order VNLE memory length $M_3$ and the resulting kernel number $N_3$ for the different approaches.

*A. Structural Complexity Reduction*

A straightforward approach to minimize the third-order kernels for a given memory length is the polynomial VNLE. This approach only uses those kernels, which are based on the cubic terms of a single sample. In this case, the number of kernels $N_3$ is equal to the third-order memory length $M_3$. The contribution of the third-order terms to the VNLE output is given by

$$y_3(k) = \sum_{k_1=0}^{M_3-1} w_3(k_1)x^3(k-k_1). \quad (2)$$

Since the polynomial VNLE skips all cross terms, a significant performance penalty compared to the full VNLE can be expected. An approach to consider more terms and therefore improve the performance compared to the polynomial VNLE is the restriction of the kernels to those, which only have two different participating samples. This means, only cubic terms as well as terms according to $x^2(k-k_n)x(k-k_m)$ are allowed, while the kernels with three different participating samples are neglected. This approach is called 2-Sam VNLE in the remainder. The contribution of the third-order terms is given by

$$y_3(k) = \sum_{k_1=0}^{M_3-1}\sum_{k_2=k_1}^{M_3-1}\sum_{k_3 \in \{k_1,k_2\}} w_3(k_1,k_2,k_3)x(k-k_1)x(k-k_2)x(k-k_3). \quad (3)$$

Note that only one kernel results for the case that the set $\{k_1,k_2\}$ contains the elements $k_1 = k_2$. This scheme contains all kernels of the polynomial VNLE and adds kernels with two different participating samples. In this case, the number of kernels is given by $N_3 = M_3^2$, resulting in a computational complexity of $C_3 = 3M_3^2$.

A third approach for a structural reduction of the kernels is based on the assumption, that those kernels with a large spacing among the participating samples are less relevant than those consisting of samples with a close spacing. For this approach, a parameter $d$ is introduced, which specifies the maximum tap spacing among samples participating in the same kernel. This approach is referred to as RI-$d$ VNLE in the following. The resulting contribution of the third-order part is given by

$$y_3(k) = \sum_{k_1=0}^{M_3-1}\sum_{k_2=k_1}^{M_d}\sum_{k_3=k_2}^{M_d} w_3(k_1,k_2,k_3)x(k-k_1)x(k-k_2)x(k-k_3), \quad (4)$$

where the upper limit is defined as $M_d = \min\{M_3-1, k_1+d-1\}$. The number of kernels for this approach can be scaled with $d$ and is given by $N_3 = (d+1)(d+2)(3M_3-2d)$. This scheme is shown for the second-order VNLE in [8-10] and referred to as diagonally-pruned VNLE.

To further reduce the number of kernels, a combination of the 2-Sam and the RI-$d$ VNLE is also possible. In this case only those kernels are considered, that have up to two participating samples with a spacing below or equal to $d$. The third-order VNLE part in this case is given by

$$y_3(k) = \sum_{k_1=0}^{M_3-1} \sum_{k_2=k_1}^{M_d} \sum_{k_3 \in \{k_1,k_2\}} w_3(k_1,k_2,k_3) x(k-k_1) x(k-k_2) x(k-k_3). \quad (5)$$

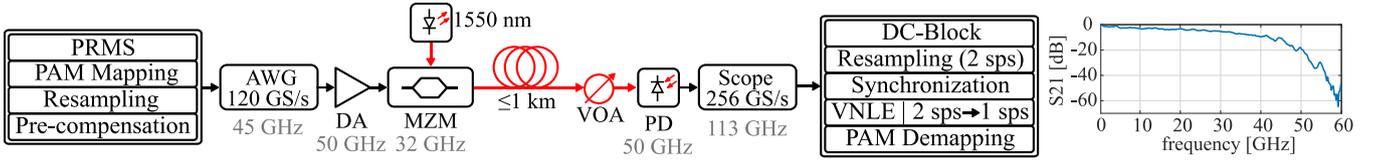

Fig. 2. Setup of the transmission system and structure of the DSP. The inset shows the spectrum of the optical back-to-back configuration.

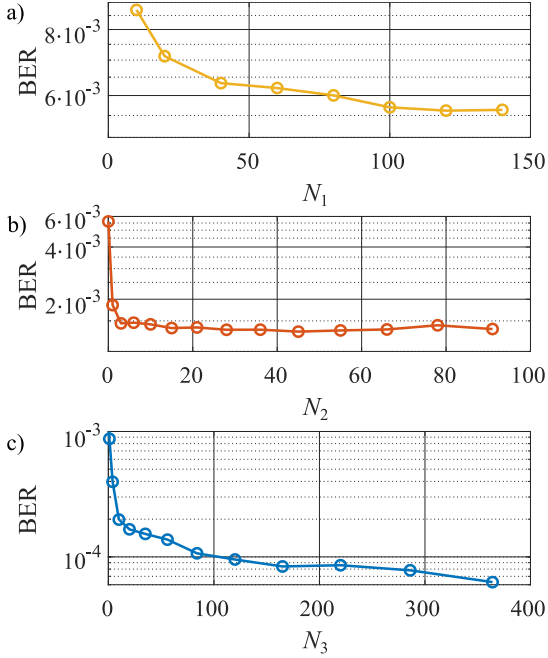

Fig. 3. Impact of the VNLE kernel number on the BER. a) shows the impact of the linear taps, b) the impact of the second-order taps and c) the impact of the third-order taps.

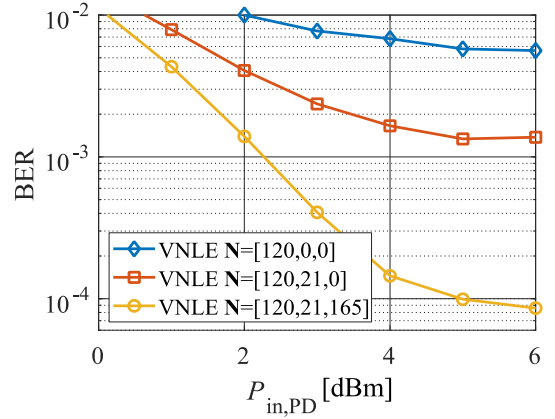

Fig. 4. Relationship of BER and input power into the PD dependent on the applied VNLE order.

The number of kernels for this combination is given by $N_3 = 2M_3 d - d(d+1) + M_3$ and the upper limit $M_d$ is defined as above.

To get an impression of the effect of the discussed kernel reduction schemes, the relationship between the memory length $M_3$ and the kernel number $N_3$ is shown in Fig. 1. The kernel number of the full third-order VNLE part increases fast with a growing memory length. Already a memory length of $M_3 = 10$ leads to more than 200 equalizer taps and therefore to a high complexity. The reduction schemes are able to reduce this number. While the polynomial VNLE has clearly the lowest complexity, the RI-$d$ is a comparably weak reduction for $d = [M_3/2]$, where [·] denotes the rounding operation. The 2-Sam approach reduces the kernel number by a factor higher than 3 for a memory length of $M_3 = 15$ and the combination of this approach with the RI-$d$ scheme leads to a further reduction. However, not the number of resulting kernels is of interest, but the tradeoff of this number and the VNLE performance. This relationship is investigated in the following transmission experiments.

## III. EXPERIMENTAL INVESTIGATIONS

The setup for the experimental investigations is shown in Fig. 2. In the transmitter DSP, a pseudo-random multilevel sequence (PRMS) is generated and mapped on the PAM symbols. The sequence is resampled to the AWG sampling rate and a pre-compensation for bandwidth limitations is performed. The digital signal is fed into the 120 GS/s AWG with a 3-dB bandwidth (BW) of 45 GHz. The resulting analog signal is amplified by a 50 GHz driver and modulated on the optical C-band carrier by a 32 GHz MZM. After transmission over the fiber, the input power into the photodiode (PD) is controlled by a variable optical attenuator (VOA). After the PD with a BW of 50 GHz, the signal is digitized at 256 GS/s by a 113 GHz oscilloscope. As last step, the receiver DSP is conducted. First, the DC-part of the signal is blocked and the sequence is resampled to 2 samples per symbol. After synchronization, the VNLE is applied. Finally, the PAM symbols are de-mapped and the BER is calculated. The inset of Fig. 2 shows the spectrum of the optical back-to-back configuration of the transmission system.

To investigate the performance of the complexity reduction approaches, we first need to find the number of VNLE kernels that is necessary for optimal performance. This is shown in Fig. 3 for the PAM6 back-to-back data. First, the linear taps are swept and the impact on the BER is observed. As visible in Fig. 3 a), a number of $N_1 = 120$ taps leads to optimal performance. Additional taps do not further improve the performance. Afterwards, this value is fixed and the number of second-order taps is optimized. A low memory length for the second order is sufficient, so that we fix it to $M_2=6$, which results in $N_2=21$ kernels, for the following investigations. As shown in Fig. 3 b), the number of second order taps can be reduced further without a strong performance penalty, so that complexity reduction schemes for the second order terms are not interesting for our setup. In other cases, the polynomial VNLE as well as the RI-$d$ VNLE scheme can be straightforwardly applied for the second order kernels, too. Fig. 3 c) shows the relationship between the third-order kernel number and the BER. The third-order kernels have a strong impact on the overall performance and a high number of kernels is necessary for optimal performance. We chose a memory length of $M_3=9$ resulting in $N_3=165$ kernels as a sufficient value.

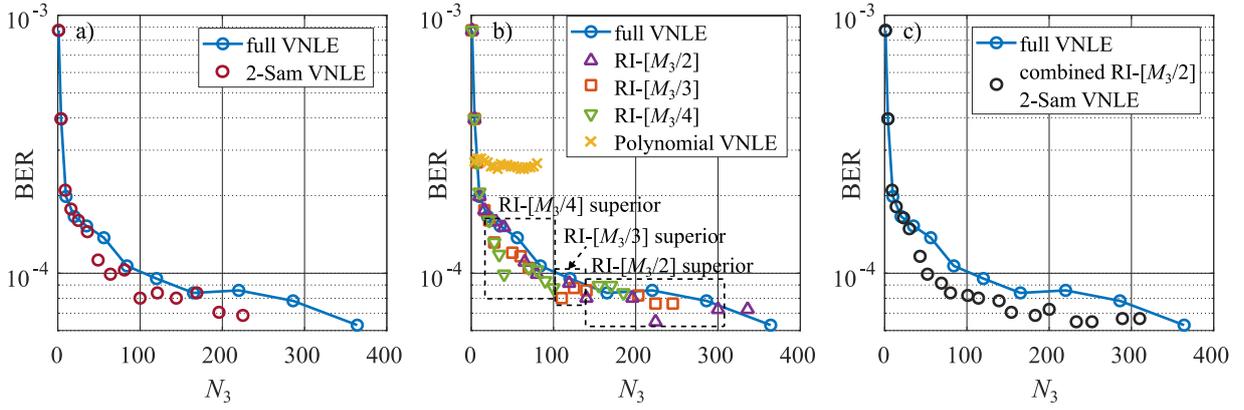

Fig. 5. Comparison of the relationship between performance and kernel number for full VNLE and complexity reduced VNLE schemes. The kernel numbers are $\mathbf{N}=[120,21,N_3]$.

Fig. 4 shows the BER over the input power into the PD for a linear FFE as well as a VNLE of second and third order. While the second-order kernels can increase the performance by a factor of 4 compared to the linear FFE, the third-order kernels lead to an additional performance improvement by a factor of more than 10. It needs to be noted that the settings in the experimental setup such as the MZM bias voltage and the peak-to-peak voltage $V_{pp}$ of the signal entering the MZM were optimized with respect to the optimal performance under application of a full third order VNLE. Therefore, the performance for the linear FFE could possibly be improved by choosing different settings. Also, the impact of the nonlinear VNLE terms changes, if different values for the bias and $V_{pp}$ are selected. However, the settings resulting in the best overall performance are of interest.

To check the effectivity of the complexity reduction schemes, we need to compare their relationship between kernel number and BER with that of the full third-order VNLE. The resulting pairs of BER and kernel number need to lie below the curve in Fig. 3 c) for an improved tradeoff. This comparison of the 2-Sam VNLE and the full third-order VNLE part is shown in Fig. 5 a). For low kernel numbers, no improvement can be realized. However, for higher numbers of third-order kernels, the 2-Sam VNLE can realize several better BER - $N_3$ pairs than the full VNLE. Based on this, we can conclude that many of the significant kernels have only one or two participating samples.

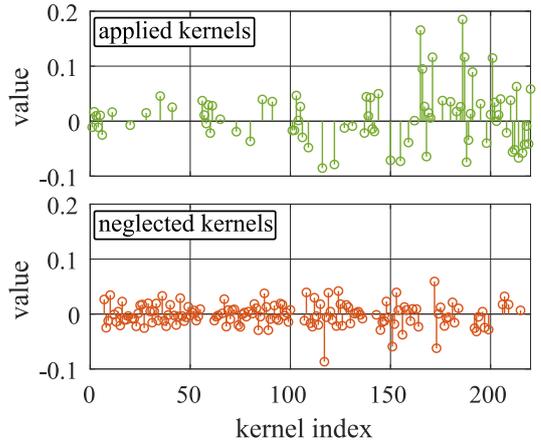

Fig. 6. Applied and neglected third-order kernels after complexity reduction by the combined RI-$d$ and 2-Sam VNLE scheme.

The same comparison can be done for the polynomial and the RI-$d$ VNLE. The results are shown in Fig. 5 b). The polynomial VNLE has a nearly constant performance for a growing kernel number. This shows, that the cubic terms alone are not sufficient for a good performance and that especially the terms that result from a large memory length are not significant. The performance of the RI-$d$ VNLE is dependent on the parameter $d$. In case it is chosen to be $d=[M_3/4]$, the kernel number is reduced significantly and a relatively good performance can be reached for kernel numbers below $N_3=100$. For slightly higher kernel numbers, a parameter of $d=[M_3/3]$ leads to good results while $d=[M_3/2]$ leads to a good tradeoff between kernel number and performance for higher kernel numbers $N_3>130$.

Finally, the combination of RI-$[M_3/2]$ and 2-Sam VNLE is investigated. The comparison between performance and kernel number between this combination and the full VNLE is shown in Fig. 5 c). The complexity reduced VNLE results in a better tradeoff between BER and third-order kernel number than the full VNLE for all investigated kernel numbers. Comparing these results with those of the individual application of both schemes in Fig. 5 a) and b) shows, that the combination leads to the best results. An example for the kernels that are applied with the combined complexity reduction scheme and the neglected kernels is shown in Fig. 6 for a memory length of $M_3 = 10$. Most of the kernels with a large value and therefore a high relevance are inside the 80 of 220 kernels that are remaining according to the specified rules. This proves, that for the used transmission system, the third-order Volterra kernels for which the tap spacing of the participating samples is small and which at the same time only have up to two different participating samples are most relevant. The kernels with three different participating samples and those with a large spacing among the samples are mainly negligible. The BER over the input power into the PD curves for the full VNLE, the combined RI- $d$ 2-Sam and polynomial VNLE are compared in Fig. 7. Compared to the full VNLE, the RI- $d$ 2-Sam VNLE can halve the number of third-order kernels without any performance degradation. Note that the reduced complexity approach uses a memory length of $M_3 = 10$ in the given example, whereas a value of $M_3 = 9$ is used for the full VNLE. The polynomial VNLE shows a worse performance than the other two schemes, but on the other hand reduces the number of third-order kernels to $N_3 = 10$. The acceptability of this performance penalty depends on the particular application.

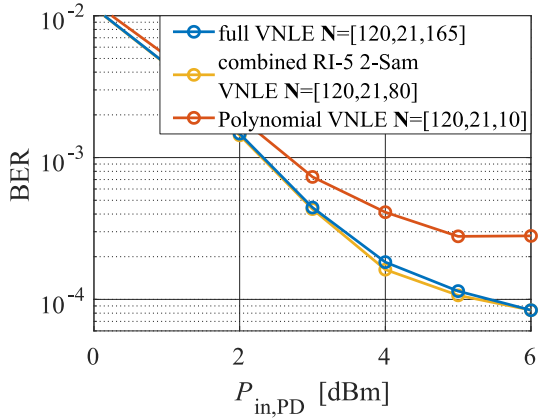

Fig. 7. Performance comparison between full VNLE, combined RI- $d$ 2-Sam and polynomial VNLE. The kernel number can be halved compared to the full VNLE without a performance penalty. The polynomial VNLE can further reduce the kernel number but shows a worse performance.

The comparison of the full VNLE and the combined RI- $d$ 2-Sam VNLE, which is shown in Fig. 5 c) for a back-to-back transmission of 90 GBd PAM6, is done for additional experimental data in Fig. 8. As well for the transmission of 90 GBd PAM6 over 1 km SSMF, as for 75 GBd PAM8 in a back-to-back configuration and over 1 km SSMF, the complexity reduced scheme shows a better performance – complexity tradeoff.

## IV. CONCLUSION

We investigated different approaches for a structural complexity reduction of the third-order VNLE for short-reach IM/DD systems. The limitation of the third-order kernels to those that only have two different participating samples as well as the limitation of the spacing between samples participating in the same kernel show a good tradeoff between performance degradation and complexity reduction compared to the full VNLE. Combining both of these rules is even more promising as shown on 90 GBd PAM6 experimental data. Generally, the complexity even with the shown structural complexity reduction approaches is relatively high, if optimal performance is approached. A reduction of the VNLE third-order kernels to numbers of $N_3 \leq 30$ cannot be realized without a performance degradation. However, the shown schemes improve the situation compared to the full VNLE.

ACKNOWLEDGEMENT

The authors thank Keysight for the loan of the M8194A AWG and the UXR1104A oscilloscope.

REFERENCES

[1] J. Wei et al.," Experimental Comparison of Modulation Formats for 200 G/λ IMDD Data Centre Networks," in *Proc. ECOC*, Tu3D.2, 2019.
[2] N. Stojanovic et al., "210/225 Gbit/s PAM-6 Transmission with BER Below KP4-FEC/EFEC and at Least 14 dB Link Budget," *in Proc. ECOC*, 2018
[3] Y. Gao et al., "112 Gb/s PAM-4 Using a Directly Modulated Laser with Linear Pre-Compensation and Nonlinear Post-Compensation," *in Proc. ECOC*, 2016.

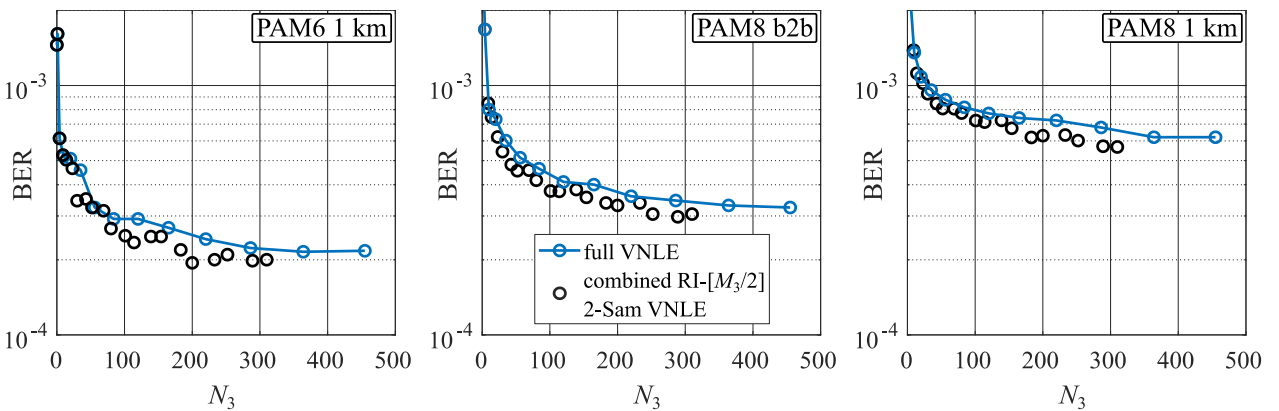

Fig. 8. Investigation of the performance – complexity tradeoff of the combined RI- $d$ 2-Sam scheme compared with the full third-order VNLE. The comparison is made for experimental data acquired in a 90 GBd PAM6 transmission over 1 km and 75 GBd PAM8 transmission in a back-to-back configuration and over 1 km SSMF in C-band. The kernel numbers are $\mathbf{N} = [120, 21, N_3]$.

[4] M. Xiang et al., "Single-Lane 145 Gbit/s IM/DD Transmission With Faster-Than-Nyquist PAM4 Signaling," *IEEE Photonics Technology Letters*, vol. 30, no. 13, pp. 1238-1241, 2018.

[5] W. Huang et al., "93% Complexity Reduction of Volterra Nonlinear Equalizer by ℓ1-Regularization for 112-Gbps PAM-4 850-nm VCSEL Optical Interconnect," in *Proc. OFC*, 2018.

[6] C. Chuang et al., "Sparse Volterra Nonlinear Equalizer by Employing Pruning Algorithm for High-Speed PAM-4 850-nm VCSEL Optical Interconnect," in *Proc. OFC*, 2019.

[7] N. Stojanovic et al., "Volterra and Wiener Equalizers for Short-Reach 100G PAM-4 Applications," *Journal of Lightwave Technology*, vol. 35, no. 21, pp. 4583-4594, 2017.

[8] E. Batista et al., "On the performance of adaptive pruned Volterra filters," *Signal Processing*, vol. 93, pp. 1909-1920, 2013.

[9] H. Xin et al., "Nonlinear Tomlinson-Harashima precoding for direct-detected double sideband PAM-4 transmission without dispersion compensation," *Opt. Express*, vol 27, pp. 19156-19167, 2019.

[10] K. Zhang et al., "Performance comparison of DML, EML and MZM in dispersion-unmanaged short reach transmissions with digital signal processing," *Opt. Express*, vol. 26, pp. 34288-24304, 2018.